\documentclass[aps, pre, twocolumn, groupedaddress, showpacs]{revtex4-1}
\usepackage{graphicx}   
\usepackage{epstopdf}
\usepackage{amssymb, amsmath, amsfonts}
\usepackage{xspace,color}

\usepackage{subfig}  

\renewcommand{\d}{{\mathrm d}}
\newcommand{\e}{{\rm e}}

\renewcommand{\hbar}{\overline h\xspace}
\renewcommand{\vr}{v_\mathrm r\xspace}

\newcommand{\eq}[1]{Eq.~#1}
\newcommand{\eqs}[1]{Eqs.~#1}
\newcommand{\fig}[1]{Fig.~#1}
\newcommand{\figs}[1]{Figs.~#1}
\newcommand{\Sec}[1]{Sec.~#1}

\newcommand{\PD}[2]{\frac{\partial #1}{\partial #2}}
\newcommand{\FD}[2]{\frac{\d #1}{\d #2}}

\begin{document}


\title{First passage times in integrate-and-fire neurons with stochastic thresholds}


\author{Wilhelm Braun}
\email[]{pmxwb1@nottingham.ac.uk}
\author{Paul C. Matthews}
\author{R\"{u}diger Thul}

\affiliation{School of Mathematical Sciences and Centre for Mathematical Medicine and Biology, University
of Nottingham, University Park, Nottingham, NG7 2RD, UK}


\date{\today}

\begin{abstract}
We consider a leaky integrate--and--fire neuron with deterministic subthreshold dynamics and a firing threshold that evolves as an Ornstein--Uhlenbeck process. The formulation of this minimal model is motivated by the experimentally observed widespread variation of neural firing thresholds. We show numerically that the mean first passage time can depend non-monotonically on the noise amplitude. For sufficiently large values of the correlation time of the stochastic threshold the mean first passage time is maximal for non-vanishing noise. We provide an explanation for this effect by analytically transforming the original model into a first passage time problem for Brownian motion. This transformation also allows for a perturbative calculation of the first passage time histograms. In turn this provides quantitative insights into the mechanisms that lead to the non-monotonic behaviour of the mean first passage time. The perturbation expansion is in excellent agreement with direct numerical simulations. The approach developed here can be applied to any deterministic subthreshold dynamics and any Gauss--Markov processes for the firing threshold. This opens up the possibility to incorporate biophysically detailed components into the subthreshold dynamics, rendering our approach a powerful framework that sits between traditional integrate-and-fire models and complex mechanistic descriptions of neural dynamics.
\end{abstract}


\maketitle


\section{Introduction}
\textit{In vivo} recordings of the membrane voltage in many types of neurons display stereotypical upstrokes known as spikes \cite{Rieke:1999wj}. The timing of these spikes has been shown to exhibit large variability. While fluctuations are often considered as obscuring biological function, the constructive role of stochasticity in neural information processing is now well established \cite{nature_review_noise, lindner_review_2004, stein_nature_review_2005}.  

The random nature of spike generation has attracted considerable attention in the field of mathematical neuroscience. Among the many models that have been suggested to describe neural stochasticity, the class of integrate-and-fire (IF) models \cite{lapicque_1907} has been used to great effect \cite{burkitt_review_1, *burkitt_review_2}. At the heart of IF models is the notion of neural excitability. A voltage spike is elicited when the membrane potential reaches a threshold. Traditionally, the threshold is considered as either constant or a given function of time that can depend on the spike history, while the membrane potential is described by stochastic differential equations \cite{gerstein_mandelbrot_1964, stein_1965, stein_1967, gs_review,lindner_schimansky_longtin_lif, touboul_faugeras_2007, chacron_lindner_longtin_2007,avila_akerberg_chacron_2011}. 

In the present study we investigate the timing of spikes that are generated when a deterministic subthreshold voltage crosses a stochastic firing threshold. A fluctuating threshold reflects experimental findings that the membrane voltage at which a spike is elicited varies. This has been demonstrated e.g. in cortical neurons and hippocampal pyramidal cells \cite{azouz_gray_2000,  henze_buzsaki_2001}. Numerous mechanisms have been suggested that could give rise to a variable threshold including adaptation, channel noise and dynamic modulation of the axon initial segment \cite{fontaine_2014,White:2000dq,Sigworth:1980ul,Grubb:2011eq}. The physiological relevance of a variable threshold was recently demonstrated in cortical neurons where synchrony detection was significantly improved \cite{azouz_gray_2000}. In \cite{coombes_thul_noisy_threshold_2011} a leaky IF model coupled to a threshold that evolves as a Gaussian process successfully described spiking behaviour of regularly firing stellate neurons within the ventral cochlear nucleus. A stochastic threshold also captures inherent uncertainty in both the detection of spiking thresholds and the spike generation mechanism. The former arises from non-standardised methods to determine threshold values from experimental records \cite{platkiewicz_brette_2010} and the fact that spikes are generated at the axon initial segment, but are often recorded at the soma. The latter results from still incomplete knowledge of the molecular components that trigger a spike \cite{Grubb:2011eq}.  

IF models with a fluctuating threshold have been investigated in the past. However, the approach presented here significantly differs from these studies.  Often a distribution of threshold values has been assumed at each time point \cite{Knight:1972ue, Barbi:2003wx,Barbi:2005ee, lindner_chacron_longtin_2005}. As such, successive threshold values are uncorrelated, and in principle arbitrarily large changes of the threshold can occur due to the unbounded support of the chosen threshold distributions. In \cite{Plesser:2000to} the concept of hazard functions is used to determine optimal parameters of the firing probability. The tested hazard functions have unbounded support, and the analysis assumes an inhomogeneous Poisson process as the spike generation mechanism \cite{Barbieri:2001bp}.  
A more general approach was taken in \cite{ttmm_2013} where a subthreshold stochastic process intersects with a frozen stochastic threshold giving rise to a phase transition in the distribution of first passage times. 

In principle it is always possible to transform an IF model with a deterministic firing threshold and stochastic subthreshold dynamics with additive noise into a model with a deterministic description of the membrane voltage and a fluctuating firing threshold (see e.g. \cite{Shimokawa:2000wz}). Indeed our model is closely related to a leaky IF model driven by white noise and subject to an exponentially decaying firing threshold  \cite{lindner_longtin_2005}. The time--dependent firing threshold can be transformed into a constant one giving rise to an exponential drive in the subthreshold dynamics \cite{lindner_longtin_2005}. Generally such explicitly time-dependent stochastic subthreshold dynamics can only be studied numerically. Perturbative expressions for the first and second moment of the first passage time (FPT) have been obtained in the limit of a weak exponential drive \cite{lindner_2004_moments_MFPT}. In contrast, we derive analytical expressions of the FPT \emph{distribution} which do not rely on a small parameter. Our findings provide quantitative insights into non-trivial effects such as the non-monotonic behaviour of the the mean FPT (MFPT) as a function of the noise amplitude.

\section{Model and governing equations}
\label{sec:model}
\begin{figure}
  \centering
 \includegraphics[width=7cm]{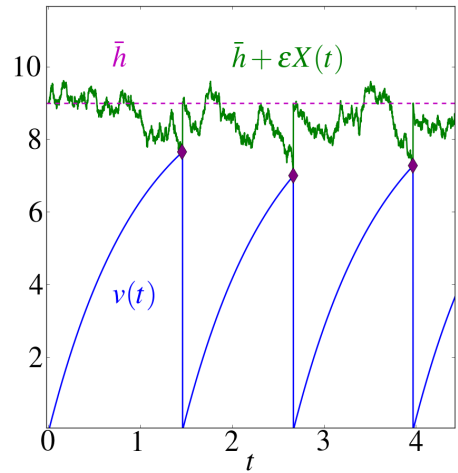}\\
  \caption{(Color online) Typical sample path of the OUP renewal model. After each spike (diamond), the OUP and the trajectory are reset to a fixed value $X(0)=v(0)=0$. Parameter values are $\beta=10,~\bar{h}=9,~\alpha=\gamma=1,~D=2,~\epsilon=1$.}
  \label{fig:fig_1}
\end{figure}

In the subthreshold regime the membrane voltage $v(t)$ obeys the ordinary differential equation of a leaky integrator
\begin{equation}
\FD v t = -\alpha v + \beta,
\label{eq:ODE_LIF}
\end{equation}
with $\alpha$ and $\beta$ positive constants. A spike occurs when $v(t)$ hits the threshold $h(t)$ from below, at which point $v$ is discontinuously reset to a constant $\vr=v(0)$, i.e.

\begin{equation}
\lim_{t \rightarrow T_n^+} v(t)=\vr\,.
\end{equation}
We hence define the firing times $T_n$ by
\begin{equation}
T_n= \inf(t | v(t) \geq h(t); t > T_{n-1})\,.
\end{equation}
The fluctuating threshold $h$ is given by
\begin{equation}
\label{eq:threshold}
h=\hbar + \varepsilon X(t)\,
\end{equation}
where $\hbar>\vr$ denotes the mean of the threshold and $\varepsilon>0$ measures the coupling strength to a stochastic process $X(t)$. We model $X(t)$ as an Ornstein--Uhlenbeck process (OUP) whose dynamics reads as \cite{Gardiner}
\begin{equation}
\label{eq:OUP}
\d X= -\gamma X \d t+ \sqrt D \d W\,.
\end{equation}
Here, $\gamma$ is the inverse of the correlation time of the OUP, $D$ is a positive constant, and $\d W$ is the increment of a standard Wiener process $W$. We choose the initial condition $X(0)=0$ such that $h(0)=\hbar$ and reset the OUP at each threshold crossing to $X(0)$. A fixed value of $X(0)$ renders the above model a \emph{renewal process}. Without the discontinuous reset of the OUP, consecutive interspike intervals (ISI) are not independent identically distributed and hence do not describe a renewal process \cite{Cox:1962tt}. \fig \ref{fig:fig_1} illustrates the generation of ISIs by the deterministic subthreshold dynamics of $v(t)$ and the OUP in the renewal regime. Note that the full non-stationary correlation function of the OUP enters our analysis, i.e.
\begin{equation}
\label{eq:OUP_corr}
 \langle X(t) X(t') \rangle = \frac{D}{2 \gamma} \left[ \e^{-\gamma|t-t'|}-\e^{-\gamma(t+t'-2T_n)} \right]\,,
\end{equation}
for $T_n\leq t,t^\prime \leq T_{n+1}$. It is worth noting that $\varepsilon$ in \eq \eqref{eq:threshold} does not need to be small. As both $v$ and $X$ are reset at a threshold crossing such that $\vr<\bar{h}$, the threshold never reaches negative values although the OUP in general takes on all values on the real line. Throughout this paper we set $\vr = 0$.

\section{Mean first passage times}
\label{sec:MFPT}
\begin{figure}[h!]
\includegraphics[scale=0.35]{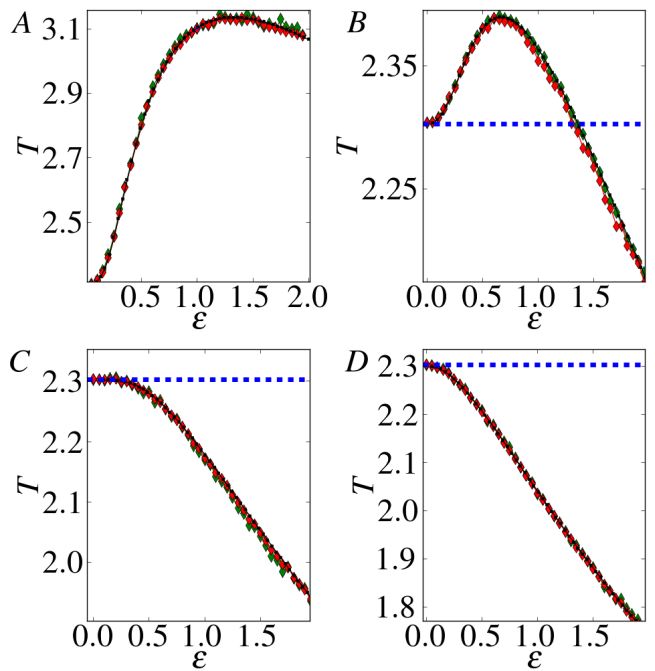}
  \caption{(Color online) MFPT $T$ as a function of $\epsilon$ for $\gamma=0.1$ (A), $0.3$ (B), $0.5$ (C) and $1.0$ (D). Threshold crossings are determined by linear interpolation (green) or using a Brownian bridge (red).  Black diamonds indicate  numerical solutions to \eq \eqref{eq:pde_mfpt}. The blue dashed line denotes the deterministic MFPT given by \eq \eqref{eq:T_det}.  Parameter values read as $\alpha = 1,~\beta=10, \bar{h} = 9,~D=2$.}
  \label{fig:brownian_boundary_mfpt_epsilon}
\end{figure}

We computed the MFPT $\langle T_{n+1} - T_{n} \rangle$ as a function of $\gamma$ and $\varepsilon$ when $\beta/\alpha>\hbar$. Angular brackets indicate the average of the firing times $T_{n+1}-T_n$ for the renewal OUP. For notational convenience, we will denote the MFPT by $T$ throughout the paper. In this case firing occurs even for a deterministic threshold at $\hbar$.  Our results are shown in \fig \ref{fig:brownian_boundary_mfpt_epsilon}. For small values of $\gamma$, the MFPT first increases as a function of $\varepsilon$ before decreasing. As we increase $\gamma$, the range over which $T$ grows monotonically as well as the amplitude both decrease. The amplitude is defined as the maximal increase of $T$ with respect to the noise--free constant firing time. The latter always exists for the parameter regime under investigation and can be computed from \eqs \eqref{eq:ODE_LIF}--\eqref{eq:threshold} with $\varepsilon=0$ as
\begin{equation}
\label{eq:T_det}
T_\mathrm{det}=\frac 1 \alpha \ln \left[\frac{\beta}{\beta-\alpha \hbar} \right] H\left ( \frac{\beta}{\alpha} -\hbar \right)\,.
\end{equation}
Here, $H$ denotes the Heaviside step function, i.e. $H(x)=1$ for $x\geq 0$ and zero otherwise. For $\gamma=0.5$ the MFPT already decays monotonically. As the OUP can be simulated exactly \cite{gillespie} and the solution for the subthreshold dynamics is known in closed form, the main error for computing ISIs results from the determination of the threshold crossing. We employed two approaches  for this. One is based on linear interpolation \cite{hansel_lif_numerical}, while the other makes use of a Brownian bridge \cite{gs_algorithm}. Both methods give almost identical results (green and red diamonds in \fig \ref{fig:brownian_boundary_mfpt_epsilon}). We used more than $10^6$ independent realisations for the computation of the MPFT at each data point. 

Another approach to compute the MFPT consists of solving the partial differential equation (PDE) 
\begin{equation}
\frac{\varepsilon^2 D}{2} \PD{^2T}{h_0^2} +\gamma(\hbar-h_0) \frac{\partial T}{\partial h_0} + \left(\beta-\alpha v_0\right) \frac{\partial T}{\partial v_0} = -1\,,
\label{eq:pde_mfpt}
\end{equation}
which describes the MFPT $T=T(v_0,h_0)$ in dependence on general initial values of the voltage and the threshold, i.e. $v(0)=v_0$ and $h(0)=h_0$. In the present study we are interested in the case $h_0 = \bar{h}$ and $v_{0} = 0$. Equation \eqref{eq:pde_mfpt} can be derived by applying the Feynman--Kac formula \cite{milstein_tretyakov_book, schuss_book} to the stochastic system given by \eqs \eqref{eq:ODE_LIF}--\eqref{eq:OUP}. In the derivation we used the stochastic differential equation for $h$, which follows from \eqs \eqref{eq:threshold}--\eqref{eq:OUP} and It\={o}'s formula \cite{Gardiner} as
\begin{equation}
\d h = -\gamma \left (h -\hbar \right) \d t + \varepsilon \sqrt D \d W\,.
\label{eq:definition_h}
\end{equation}
We solve \eq \eqref{eq:pde_mfpt} on a two-dimensional rectangular domain bounded by the line $v_0=h_0$, on which $T=0$. We use no-flux boundary conditions on the remaining three sides  of the rectangle. The size of the rectangle is chosen such that these three sides are sufficiently far away from the point of interest $(\vr,\hbar)$ eliminating any impact of the no-flux boundary conditions on the solution $T(\vr,\hbar)$ \cite{tuckwell_2}.

\begin{figure}[h!]
\includegraphics[width=7cm]{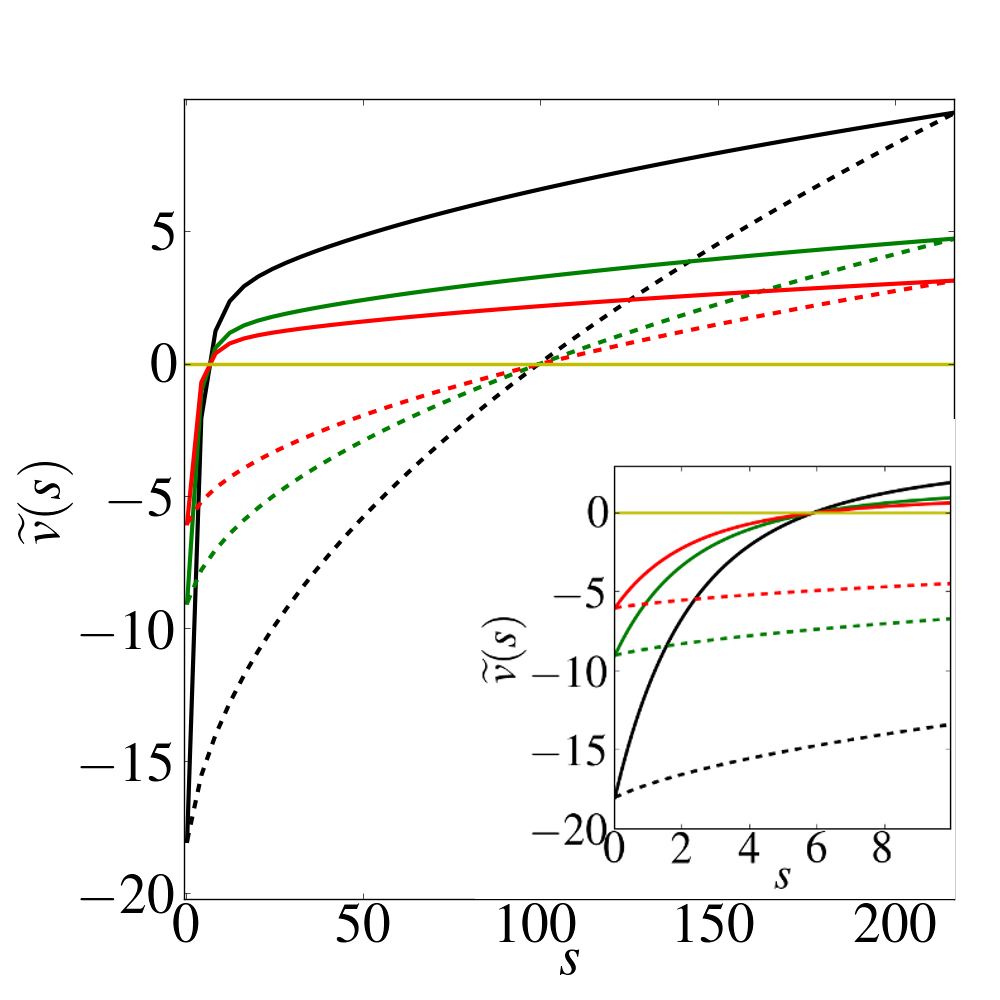}
\caption{(Color online) Subthreshold voltage $\widetilde v(s)$ for different values of $\gamma$ and $\varepsilon$. Dashed lines correspond to $\gamma=1$ and solid lines to $\gamma =0.1$ for $\varepsilon=0.5$ (black), $1$ (green) and $1.5$ (red). The inset shows a blow-up for small times $s$. Parameter values are $h(0) = \bar{h},~\alpha=1$, $\beta=10$, $\hbar=9$ and $D=2$.}
\label{fig:tranformed_boundary}
\end{figure}

Insights into the non-monotonic behaviour of the MFPT can be obtained by introducing a new stochastic variable $g=\e^{\gamma t} X$, where the dynamics for $X$ is given by \eq \eqref{eq:OUP}. Using It\={o}'s formula, the dynamics of $g$ obeys
\begin{equation}
\d g= \sqrt D \e^{\gamma t} \d W\,,
\end{equation}
and the time-dependent variance of $g$ follows as 
\begin{equation}
\label{eq:var_g}
s(t)= \langle g^2(t) \rangle = \frac{D}{2\gamma} \left( \e^{2\gamma t}-1 \right)\,.
\end{equation}
We can interpret \eq \eqref{eq:var_g} as the definition of a new time $s$ in terms of the original time $t$. We then define a new stochastic process $V(s)=g(t(s))-g(0)$ in time $s$ where $t(s)$ denotes the inverse of $s(t)$, which always exists since $s(t)$ is strictly monotonically increasing. The variance of $V$ is $\langle V^2(s) \rangle-\langle V(s) \rangle^2=s$, and together with $V(0)=0$ this renders $V$ a standard Brownian motion. This transformation is discussed extensively in \cite{mehr_mcfadden_1965} and was first used in \cite{doob_1949}. From the threshold crossing condition in the new time $s$, $v(t(s))=h(t(s)) = \hbar + \varepsilon X(t(s))$,  we find
\begin{equation}
V(s)=\left(\frac{v(t(s))-\hbar}{\epsilon}\right) \e^{\gamma t(s)}-X(0)=\widetilde v(s)\,,
\end{equation}
where we have used
\begin{equation}
X(s)=\e^{-\gamma t(s)} \left(V(s)+X(0)\right) \,,
\end{equation}
which follows from the definitions of $g$ and $V$. Therefore, the crossing of a voltage $v(t)$ through a threshold $h(t)$ that is described by an OUP is equivalent to the crossing of a new subthreshold voltage $\widetilde v(s)$ through a standard Brownian motion $V(s)$.  Figure \ref{fig:tranformed_boundary} shows the new subthrehold voltage $\widetilde v(s)$ for different values of $\gamma$ and $\varepsilon$.
We observe that for a fixed value of $\gamma$ the zero crossing of $\widetilde v$ is independent of $\varepsilon$. It can be shown that this crossing occurs at a value $s_0$ that corresponds to the deterministic first passage time (\eq \eqref{eq:T_det}). Note that the value of $s_0$ grows as we increase $\gamma$. For a fixed $s<s_0$ the value of $\widetilde v(s)$ increases as we increase $\varepsilon$. Therefore a given realisation of the standard Brownian motion $V$ will intersect with the subthreshold voltage $\widetilde v$ earlier for larger $\varepsilon$ than for smaller $\varepsilon$ for times $s<s_0$. For $\gamma=1$ almost all crossings of $\widetilde v$ and $V$ occur before $s_0$, which is reflected in the monotonically decreasing behaviour of the MFPT $T$ as a function of $\varepsilon$ (\fig \ref{fig:brownian_boundary_mfpt_epsilon}). For smaller values of $\gamma$ there is a significant probability that no threshold crossing occurs before $s_0$. Since $\widetilde v(s)$ is a decreasing function of $\varepsilon$ for fixed $s>s_0$, a given realisation of $V$ that has not intersected with $\widetilde v$ before $s_0$ will cross $\widetilde v$ earlier for smaller values of $\epsilon$ than for larger values. This realisation of $V$ has a larger FPT for growing values of $\varepsilon$. The above arguments hold for single realisations of $V$ and hence individual FPT. To elucidate the increase of the MFPT for growing $\varepsilon$ we need to differentiate between different processes. On the one hand more realisations of $V$ could intersect with $\widetilde v$ at times larger than $s_0$. On the other hand the spread of FPTs that are larger than $s_0$ grows. To determine which factor --- or the combination of both --- is relevant for the current observations we compute numerically the FPT distributions.

\section{Distribution of first passage times}
\label{sec:distribution_mfpt}
\begin{figure}[h!]
\includegraphics[scale=0.35]{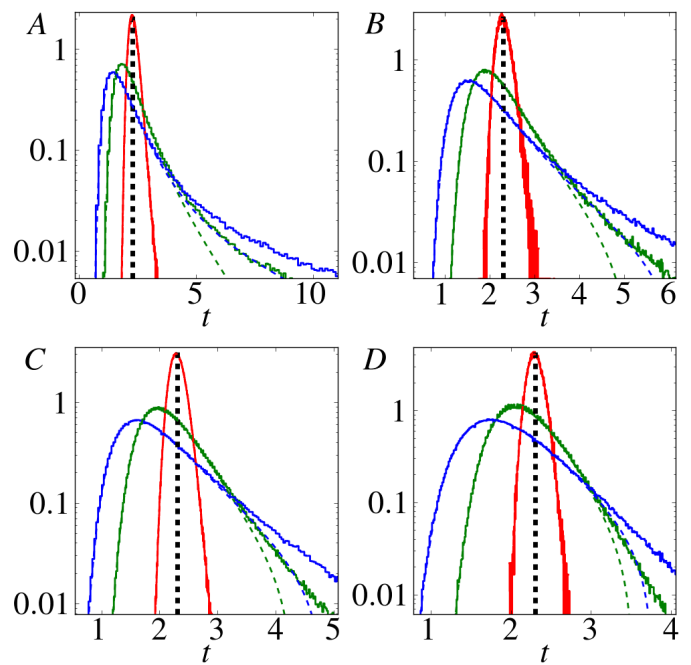}
  \caption{(Color online) FPT histograms for $\gamma=0.1$ (A), $0.3$ (B), $0.5$ (C) and $1.0$ (D) and  $\epsilon=0.1$ (red), $0.5$ (green) and $1$ (blue). The black lines indicates the deterministic FPT. The dashed lines correspond to $\mathcal F^{2}(t)$, while the solid lines result from direct numerical simulations. Parameter values are $\alpha = 1.0$, $\beta=10$, $\bar{h}=9$, $D=2$. }
  \label{fig:brownian_boundary_histograms}
\end{figure}

\begin{figure}[h!]
\includegraphics[width=7cm]{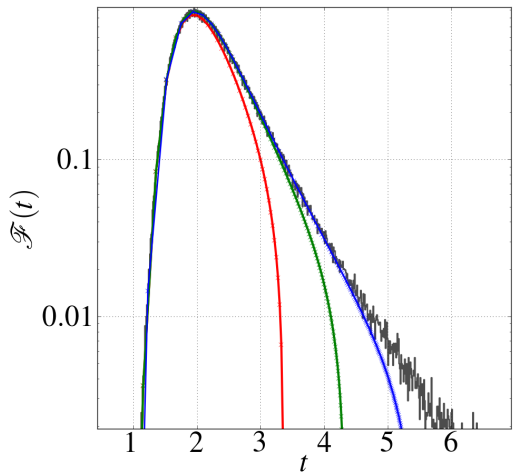}
  \caption{(Color online) FPT distributions $\mathcal F^k(t)$ for $k=1$ (red), $2$ (green) and $3$ (blue). The grey line corresponds to results from direct numerical simulations.  Parameter values are $\gamma =0.5,~\epsilon=0.5$, $D=2$, $\bar{h}=9,~\beta =10.$}
  \label{fig:brownian_boundary_3_terms_exploratory}
\end{figure}

Our calculation of FPT distributions is based on an approach developed by Durbin and Williams \cite{durbin_williams_1992}. They derived an alternating series for the FPT density $p$ of a Brownian motion through a curved boundary. Using the notation of \Sec \ref{sec:MFPT} for the transformed dynamics in time $s$, we have 
\begin{equation}
\label{eq:DW}
p(s)=\sum_{i=1}^k (-1)^{i-1}q_i(s)+(-1)^kr_k(s)\,,
\end{equation}
with
\begin{equation}
q_i(s)=\int_0^s \int_0^{s_1} \cdots \int_0^{s_{i-2}} K_i(s) f_i(s) \d s_{i-1}\cdots \d s_1\,.
\end{equation}
Here, we introduced
\begin{equation}
\label{eq:DW-Ki}
K_i(s)= \prod_{j=1}^i \left[\FD{\widetilde v(s_{j-1})}{s}- \frac{\widetilde v(s_{j-1})-\widetilde v(s_j)}{s_{j-1}-s_j}\right],
\end{equation}
where we formally set $\widetilde v(s_i)=s_i=0$. $f_i(s)$ denotes the joint probability density of $V(s)$, $V(s_1),\ldots,V(s_{i-1})$ at the boundary values $\widetilde v(s)$, $\widetilde v(s_1),\dots,\widetilde v(s_{i-1})$. In \eq \eqref{eq:DW} the term $r_k(s)$ represents the error that is made by truncating the infinite series after $k$ terms. Note that \eq \eqref{eq:DW-Ki} differs from the original expression since we study down- crossings of a Brownian motion instead of up-crossings as in \cite{durbin_williams_1992}. For practical purposes it is convenient to introduce ${\cal F}^k(s)=\sum_{i=1}^k (-1)^{i-1}q_i(s)$. We can transform results obtained in the new time $s$ to the original time $t$ via  ${\cal F}^k(t)={\cal F}^k(s(t)) \d s(t)/\d t $. In \fig \ref{fig:brownian_boundary_histograms} we compare ${\cal F}^2(t)$ with direct numerical simulations for different values of $\gamma$ and $\varepsilon$. For $\gamma=0.1$ we observe that for increasing $\varepsilon$ the maximum of the distribution slightly shifts to the left, while the tail of the distribution becomes significantly longer. This wider spread of long FPTs is the reason for the non-monotonic behaviour of the MFPT as it dominates the increase of trajectories with shorter FPTs.
For larger values of $\gamma$ the tails of the distributions still grow as we increase $\varepsilon$, but the shift of the distribution to shorter times is more pronounced. The latter leads to the monotonic decrease of the MFPT for increasing $\varepsilon$. The occurrence of long FPTs for small values of $\gamma$ can be attributed to the correlation time of the OUP $\tau_\mathrm{OU}=1/\gamma$. For small values of $\gamma$, changes in the OUP occur slowly. Since for times larger than $1/\alpha$, the voltage has almost reached its equilibrium $\beta/\alpha$, the OUP can then spend a considerable time in the vicinity of the voltage without crossing it. It is worth noting that $\mathcal F^{2}(t)$ captures the full FPT distribution extremely well for most of the probability mass. $\mathcal F^{2}(t)$ only fails at correctly predicting the tail of the FPT distribution. The extent to which $\mathcal F^k(t)$ agrees with the true distribution depends on the $k$. As we illustrate in \fig \ref{fig:brownian_boundary_3_terms_exploratory} $\mathcal F^k(t)$ approximates results from direct simulations better as we increase $k$. 
In their original contribution Durbin and Williams prove convergence of \eq \eqref{eq:DW} under the assumption that the intercept at the origin of the tangent line of $\widetilde v(s)$ for any $s$, i.e. $\widetilde v(s)/s-\d \widetilde v(s)/\d s$, does not change sign. Inspection of \fig \ref{fig:tranformed_boundary} shows, however, that the intercept is negative for small values of $s$ and positive for large values. Our results suggest that convergence of the FPT distribution also occurs if the above assumption is violated, but a rigorous proof is still missing.

\section{Cumulative density functions}
\begin{figure}[h!]
\includegraphics[width=7cm]{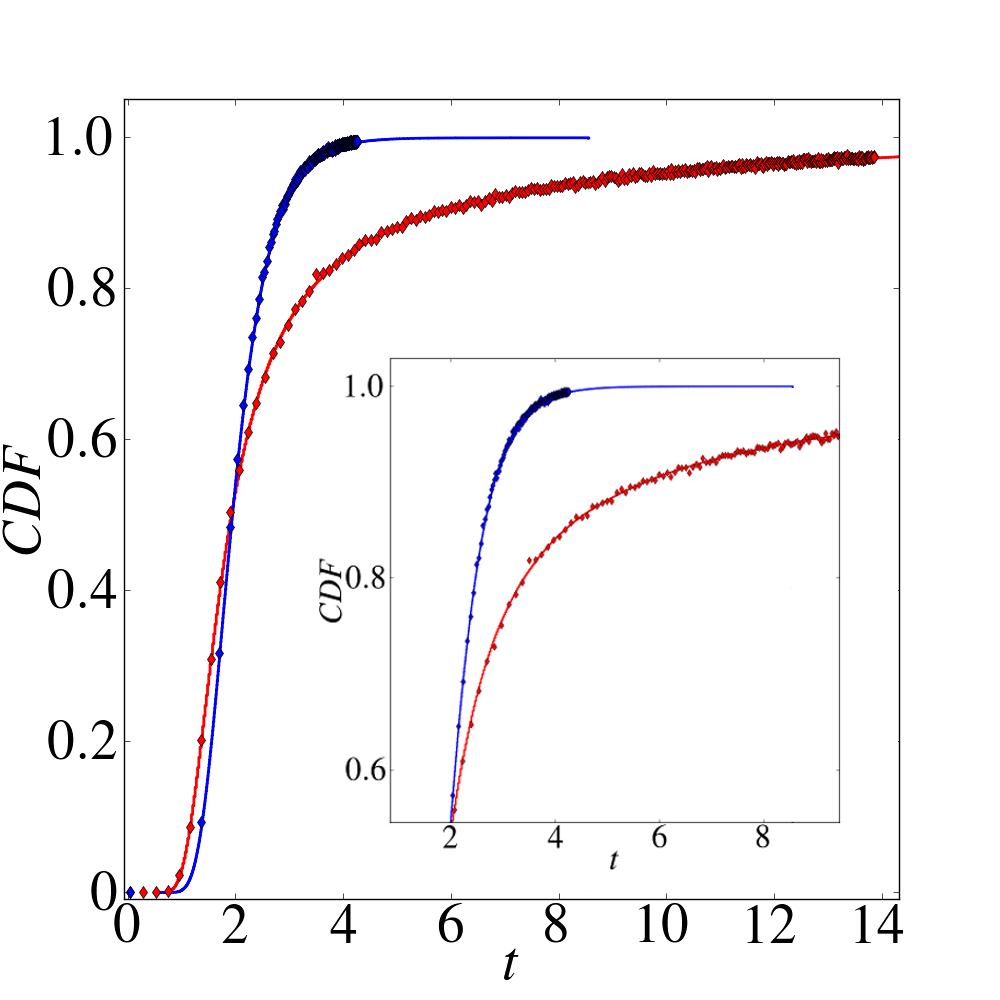}
  \caption{(Color online) Cumulative distribution function (CDF) of FPTs for $\gamma=0.1$ (red) and $\gamma=1.0$ (blue). Diamonds represent results using the quasi-analytical approach in \cite{wang_poetzelberger_1997}, while solid lines correspond to results from direct numerical simulations. Parameter values are $\alpha=1,~\epsilon=1.0,~D=2,~\bar{h}=9,~\beta=10$.}
  \label{fig:CDF}
\end{figure}

We next studied the emergence of long tails in the FPT distributions in more detail. In \fig \ref{fig:CDF} we show the cumulative distribution function (CDF) of the FPTs for different values of $\gamma$. We employed the method developed by Wang and P\"{o}tzelberger \cite{wang_poetzelberger_1997}. The CDF is expressed as an expectation value  $\langle g \rangle$ using a piecewise linear approximation of the boundary $\widetilde v$.  An advantage of this method is that each instance of $g$ is known in closed form and only the average needs to be computed numerically.  We observe that for times smaller than 2 the two CDFs agree well. 
This entails that the FPT distributions almost coincide since they follow from the corresponding CDFs by differentiation with respect to time. The CDF for $\gamma=1$ levels off at significantly earlier times than the CDF for $\gamma=0.1$ illustrating the more pronounced tail of the FPT distribution for $\gamma=0.1$. For comparison we also include results from direct numerical simulations, which lie on top of the quasi-analytical results for the CDF. The findings in \fig \ref{fig:CDF} provide an independent test for the non-monotonic behaviour of the MFPT. Moreover, convergence of the CDF as we increase the number of linear segments to approximate $\widetilde v$ is guaranteed. 

\begin{figure}[h!]
\includegraphics[width=7cm]{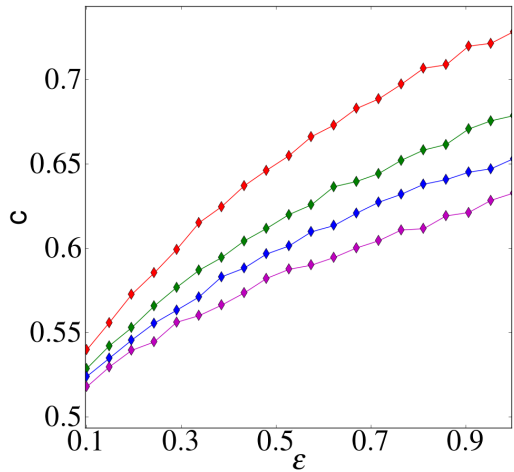}
  \caption{(Color online) Probability $c$ computed using \cite{wang_poetzelberger_1997} that $V(s)$ has crossed $\widetilde v(s)$ at least once in the time interval $(0,s_{0}]$ as a function of $\epsilon$ for $\gamma=1$ (red), $0.5$ (green), $0.3$ (blue) and $0.1$ (purple). Parameter values are $\alpha=1,~D=2,~\bar{h}=9,~\beta=10$.}
  \label{fig:probc}
\end{figure}

The shift of the FPT distributions to smaller times can also be understood by computing the probability $c$ that a given realisation of $V(s)$ intersects with $\widetilde v(s)$ at times $s<s_0$.  We computed $c$ based on the approach in \cite{wang_poetzelberger_1997} and plot results as a function of $\varepsilon$ for different values of $\gamma$ in \fig  \ref{fig:probc}. We observe that $c$ is an increasing function of $\varepsilon$. Hence a growing number of realisations of $V$ have FPTs that are smaller than $s_0$. As we increase $\gamma$ for fixed $\varepsilon$ we find that $c$ grows monotonically demonstrating that more realisations of $V$ reach $\widetilde v$ before $s_0$. The rate of increase of $c$ with $\varepsilon$ is more pronounced for larger values of $\gamma$. Therefore, the relative increase in realisations of $V$ that have FPTs smaller than $s_0$ is stronger for larger values of $\gamma$ compared with smaller values. Taken together the results in \figs \ref{fig:CDF} and \ref{fig:probc} demonstrate the subtle interplay between the emergence of long tails in the FPT distribution and the increase in the probability mass of FPTs smaller than $T_\mathrm{det}$ in the generation of non-monotonic MFPTs.

\section{Discussion}

In this paper, we have formulated and investigated an IF model for neural firing with a stochastic process controlling the firing threshold, motivated by experimental studies showing apparent variability or randomness in the spiking threshold. The model is designed to be as simple as possible while allowing fluctuations of the threshold around a 'preferred' value $\bar h$. We therefore use an OUP reverting to a mean $\bar h$ for the threshold, and a leaky IF model for the sub-threshold dynamics $v$. The simplicity of the model permits the use of techniques from stochastic calculus \cite{milstein_tretyakov_book}.

We have demonstrated that the MFPT may depend non-monotonically on the noise amplitude $\epsilon$, finding a maximum of the MFPT as the noise strength is increased. The occurrence of a minimum in spiking activity at non-vanishing noise strength has been termed \textit{inverse stochastic resonance} and has been described in stochastic Hodgkin--Huxley models \cite{gutkin_jost_tuckwell_2009}. The emergence of inverse stochastic resonance in these models is directly linked to the phase-space structure of the corresponding deterministic models. In contrast the deterministic LIF model does not predict the complex behaviour of its stochastic counterpart. Inverse stochastic resonance has been observed \textit{in vivo} \cite{paydarfar_forger_clay_2006} and may even have therapeutic applications \cite{bloch_salisbury_2009}. 

Local extrema of the MFPT have been reported before. In \cite{lindner_2004_moments_MFPT} escape from a linear and parabolic potential in the presence of a weak exponential driving force is considered, while a Kramers problem with a fluctuating barrier is investigated in \cite{iwaniszewski_1996}. In both studies  extrema of the MFPT emerge when a time scale is varied, which is the decay rate of the weak exponential forcing in \cite{lindner_2004_moments_MFPT}  and the correlation time of the random barrier in  \cite{iwaniszewski_1996}. The dependence of the MFPT as a function of the noise amplitude is always monotonic in these studies. This sets our results apart from these earlier findings. Moreover, transforming our model to one with a constant threshold and an exponential drive in the subthreshold voltage results in an amplitude of the time-dependent drive that is not small for the parameter values that we have considered. This  precludes us from applying the perturbative results in \cite{lindner_2004_moments_MFPT,lindner_longtin_2005} for computing the first two moments of the MFPT. An exponential drive of sufficient amplitude and decay time is essential for observing the non-monotonic behaviour of the MFPT. The non-monotonicity results from the competition between small passage times, which arise when the OUP crosses $v$ during the rising phase of the subthreshold voltage, and large passage times, which occur when the OUP intersects with the almost stationary value of $v$.

The dependence of the MFPT on the noise strength was found both by direct Monte Carlo simulations and by solving a backward Kolmogorov-type PDE, with virtually identical results. The increase of the MFPT with noise strength can be explained quantitatively by inspection of the FPT distributions. As the noise becomes stronger, the peak of the FPT distribution moves towards shorter firing times, but the tail of the distribution strengthens significantly, so that rare long times between firings become more frequent. Further understanding of this effect was achieved by transforming the dynamics of the system to that of a standard Brownian motion. This allows a series method \cite{durbin_williams_1992} to be used to obtain approximations to  the FPT distribution analytically; this series appears to converge rapidly, showing good agreement with numerical results.

A number of previous studies on LIF models with constant threshold employed different noise intensities of the OUP. For example, a noise scaling of $\sqrt{2 D\gamma}$ was used in \cite{middleton_pre_2003}, which entails that the variance of the OUP is given by $D$ and hence does not depend on the correlation time $1/\gamma$. By setting the noise intensity to $\sqrt{2 D}\gamma$ the authors of \cite{Schwalger_Schimansky_2008} ensure that the variance of the subthreshold voltage remains almost constant in the limit of long correlation times of the OUP, for which the variance is given by $D\gamma$. Since the emergence of the maximum of the MFPT depends on the interplay of long correlation times and sufficient variance of the OUP we tested both of the above noise intensities. Our results show that the new noise scalings of the OUP have no effect on the non-monotonicity of the MFPT. The maximum still exists but is shifted to larger values of $\varepsilon$ for a given value of $\gamma$. 

An interesting question is how broadly applicable the transformation of a stochastic threshold to Brownian motion is, particularly since an IF model with additive noise subthreshold dynamics and constant firing threshold can always be transformed into a model where the subthreshold voltage evolves deterministically and hits a stochastic threshold. If the stochastic threshold is given by a Gauss--Markov process a transformation to Brownian motion always exists \cite{mehr_mcfadden_1965}.  A wider class of practically relevant stochastic threshold processes are general diffusion processes \cite{Gardiner}. Here a transformation to Brownian motion is only feasible for specific forms of the drift and diffusion functions. Wang and P\"otzelberger  \cite{wang_poetzelberger_2007} provide a practical PDE that these functions need to satisfy to derive an equivalent Brownian motion. For the most general conditions to recast a diffusion process as Brownian motion we refer the reader to \cite{Bluman:1980td}.

Future work will consider two variations of the set-up of the problem investigated here. We have considered a `renewal' model in which the threshold is reset to a constant value after firing, but an alternative procedure is to set it to a value selected from the statistical distribution of the OUP. A second possibility is to use a full spike train, in which the threshold OUP is not reset after each spike, but evolves freely. This will give rise to serial correlations of the interspike intervals, and it will be of interest to investigate  whether the serial correlations display stronger or weaker correlations than the OUP threshold process itself.

\begin{acknowledgments}
We would like to thank Michael Tretyakov and the anonymous reviewers for very helpful comments. WB was supported by a Vice-Chancellor's Scholarship for Research Excellence at the University of Nottingham. 
\end{acknowledgments}

\bibliography{literature_paper.bib}

\end{document}